# Two-Dimensional Bipolar Magnetic Semiconductor with High Curie Temperature and Electrically Controllable Spin Polarization Realized in Exfoliated Cr(pyrazine)$_2$ Monolayer

Xiangyang Li,[†,§,Δ] Haifeng Lv,[†,§,Δ] Xiaofeng Liu,[†,§] Xiaojun Wu,[*,†,§] Xingxing Li,[*,†,§] and Jinlong Yang[*,†,‡,§]

[†]Synergetic Innovation Center of Quantum Information and Quantum Physics, University of Science and Technology of China, Hefei, Anhui 230026, China
[‡]Hefei National Laboratory for Physical Sciences at the Microscale, University of Science and Technology of China, Hefei, Anhui 230026, China
[§]Department of Chemical Physics, University of Science and Technology of China, Hefei, Anhui 230026, China

**ABSTRACT:** Exploring two-dimensional (2D) magnetic semiconductors with room temperature magnetic ordering and electrically controllable spin polarization is a highly desirable but challenging task for nanospintronics. Here, through first principles calculations, we propose to realize such a material by exfoliating the recently synthesized organometallic layered crystal Li$_{0.7}$[Cr(pyz)$_2$]Cl$_{0.7}$·0.25(THF) (pyz = pyrazine, THF = tetrahydrofuran) [Science 370, 587 (2020)]. The feasibility of exfoliation is confirmed by the rather low exfoliation energy of 0.27 J/m$^2$, even smaller than that of graphite. In exfoliated Cr(pyz)$_2$ monolayer, each pyrazine ring grabs one electron from the Cr atom to become a radical anion, then a strong d-p direct exchange magnetic interaction emerges between Cr cations and pyrazine radicals, resulting in room temperature ferrimagnetism with a Curie temperature of 342 K. Moreover, Cr(pyz)$_2$ monolayer is revealed to be an intrinsic bipolar magnetic semiconductor where electrical doping can induce half-metallic conduction with controllable spin-polarization direction.

## INTRODUCTION

Two-dimensional (2D) intrinsic ferromagnetic semiconductors, integrating semiconductivity, ferromagnetism, and low dimensionality, open up exciting opportunities for nanoscale spintronic devices. Unfortunately, the usually weak ferromagnetic superexchange magnetic interaction makes their Curie temperature ($T_c$) far below room temperature, greatly limiting their practical applications. So far, the experimentally realized 2D ferromagnetic semiconductors, CrX$_3$ (X = Cl, Br, I)[1, 2] and Cr$_2$Ge$_2$Te$_6$,[3] only retain their ferromagnetism below about 45 K. Theoretically, while a number of 2D ferromagnetic semiconductors have been predicted, such as CrOCl,[4] CrSBr,[5] GdI$_2$,[6] CrWI$_6$,[7] CrWGe$_2$Te$_6$[7] and K$_3$Co$_2$[PcCoO$_8$],[8] most of $T_c$ values are still much lower than room temperature with a few exceptions, e.g., CrSeBr ($T_c$ =500 K),[9] Fe$_3$P ($T_c$ =420 K)[10] and CrMoS$_2$Br$_2$ ($T_c$ =360 K).[11] It still remains a challenge to search for 2D high-$T_c$ ferromagnetic semiconductors.

To solve the above issue, we have theoretically proposed a general scheme to realize room temperature magnetic semiconductors in 2D metal organic frameworks (MOFs) by introducing strong d-p direct ferrimagnetic exchange interactions between transition metal cations and magnetic organic linker radical aions.[12, 13] The applicable organic linkers are conjugated electron acceptors such as pentalene ($T_c$ =560 K) and diketopyrrolopyrrole (DPP) derivatives ($T_c$ =316 K). Inheriting the high tunability of MOFs,[14] the electrical and magnetic properties of such organometallic ferrimagnetic semiconductors are expected to be easily enriched and modulated by changing the transition metals, organic linkers, or framework geometries, thus enabling their versatile applications in the field of emergent magnetoelectronic, magnetic sensing, and recording technologies.[15-19] Besides, it would also be fairly easy to integrate other functions, such as ferroelectricity,[8] ferroelasticity,[20] quantum topology,[21] photoelectricity[22] and chirality,[23, 24] to prepare multifunctional magnetic semiconductors. Despite the great potentials, the development of 2D organometallic ferrimagnetic semiconductors is still at an early stage and experimental realization keeps an open question.

In addition to obtaining high-$T_c$, it is also demanded to develop functional magnetic semiconductors with electrically controllable spin polarization. For this purpose, bipolar magnetic semiconductors (BMS),[25] characterized by oppositely and fully spin polarized valence and conduction band edges, have been theoretically proposed, which can provide completely spin polarized currents with the spin polarization direction reversible by altering the polarity of applied voltage gate.[26,27] Up to now, a number of BMS materials have been theoretically designed,[6, 28-34] but they are difficult to realize in experiment, because most of them are extrinsic and require precisely controlled chemical or physical modifications,[28, 29] or their magnetic orders are only stable at very low temperatures.[6, 31-35] Therefore, it is urgent to find 2D BMS materials with both intrinsic magnetism and high temperature magnetic stability.

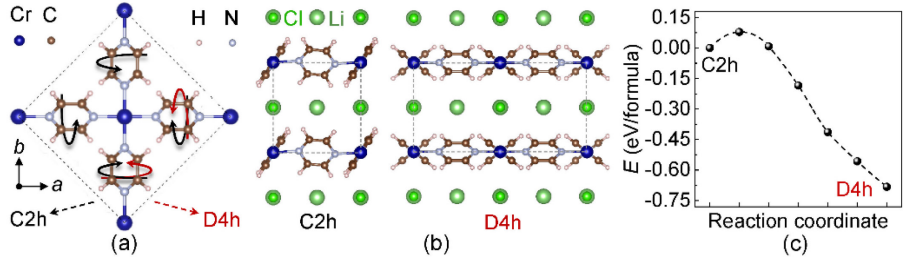

**Figure 1.** (a) Schematic representation of the two rotational isomers (C2h and D4h symmetry) of Li$_{0.7}$[Cr(pyz)$_2$]Cl$_{0.7}$·0.25(THF) (pyz = pyrazine, THF = tetrahydrofuran). For C2h crystal, the plane of pyz rings at the para positions are inclined to the same side by about 45°, as indicated by the black arrows. For D4h crystal, all the pyz rings around Cr ions are arranged obliquely clockwise or counterclockwise, as indicated by the red arrows. (b) Side view of optimized structures of C2h and D4h crystals. (c) Energy profile of transition from C2h to D4h structure.

Here, based on first principles calculations, we propose that 2D organometallic ferrimagnetic semiconductor, i.e. Cr(pyz)$_2$ monolayer, with high Curie temperature and electrically controllable spin polarization is experimentally accessible by exfoliating existing layered crystal Li$_{0.7}$[Cr(pyz)$_2$]Cl$_{0.7}$·0.25(THF) (pyz = pyrazine, THF = tetrahydrofuran).[15] The high Curie temperature (~ 342 K) stems from the strong $d$-$p$ direct ferrimagnetic exchange interaction between Cr and magnetic pyz. The electrical control of spin polarization originates in 2D Cr(pyz)$_2$'s intrinsic BMS property, which can present half-metallicity with spin-polarization direction tuned by the type of electrical doping.

## RESULTS AND DISCUSSION

The bulk crystal Li$_{0.7}$[Cr(pyz)$_2$]Cl$_{0.7}$·0.25(THF) features a layered structure composed of alternating stacks of charge-neutral Cr(pyz)$_2$ and Li$_{0.7}$Cl$_{0.7}$ layers (Figure 1). For the Cr(pyz)$_2$ layer, each Cr$^{2+}$ ion is coordinated by four pyz units in a nearly square planar coordination pattern, and each pyz unit is coordinated by two adjacent Cr ions with a linear coordination geometry. Because the dihedral angle $\theta$ between the pyz and $ab$ lattice planes has two possibilities (+45°, −45°), two stable rotational isomers (C2h and D4h symmetry) exist, and the exact crystal structure has not been determined in experiment. Figure 1(a) shows the detailed rotation operations to obtain the two isomers. For C2h crystal [Figure 1(b)], the plane of pyz rings at the para positions are inclined to the same side by about 45°. For D4h crystal [Figure 1(b)], all the pyz rings around Cr ions are arranged obliquely clockwise or counterclockwise at the same time. The C2h structure can be easily transformed to the D4h structure with a small transition barrier of 0.08 eV per formula and an energy release of 0.69 eV per formula [Figure 1(c)]. Note that the D4h structure is very similar to those of our previously proposed Cr(pentalene)$_2$[12] and Cr(DPP)$_2$.[13] Above all, we theoretically identify the D4h crystal as the most energetically favorable structure. Therefore, the D4h structure is used in the following studies.

Due to the relatively big interlayer distance (3.60 Å) between charge-neutral Cr(pyz)$_2$ and Li$_{0.7}$Cl$_{0.7}$ layers, the interlayer interaction is expected to be weak. Thus, a 2D Cr(pyz)$_2$ monolayer may be attained through the mechanical cleavage method. As shown in Figure 2(a), the predicted exfoliation energy is only 0.27 J/m$^2$, which is even smaller than that of graphite (0.37 J/m$^2$),[36] directly indicating the monolayer can be easily prepared from the bulk crystal.

The stability of exfoliated 2D Cr(pyz)$_2$ monolayer is further assessed by its phonon band structure and molecular dynamics simulation. In the calculated phonon dispersion curves [Figure 2(b)], no obvious imaginary frequency is observed, indicating that 2D Cr(pyz)$_2$ monolayer is dynamically stable. The rather big lattice constant results in the dispersionlessness of phonon bands, while the existence of a large amount of soft phonon modes means the flexibility of Cr(pyz)$_2$ monolayer.[12] To evaluate the thermal stability, ab initio molecular dynamics (AIMD) simulation at 400 K lasting for 9 *ps* is performed [Figure 2(c)]. The evolution of energy with time fluctuates near its equilibrium value without sudden drop. The structure is maintained well without any reconstruction after 9 *ps*, as shown in the insets of Figure 2(c). Thus, 2D Cr(pyz)$_2$ monolayer is thermally stable at room temperature.

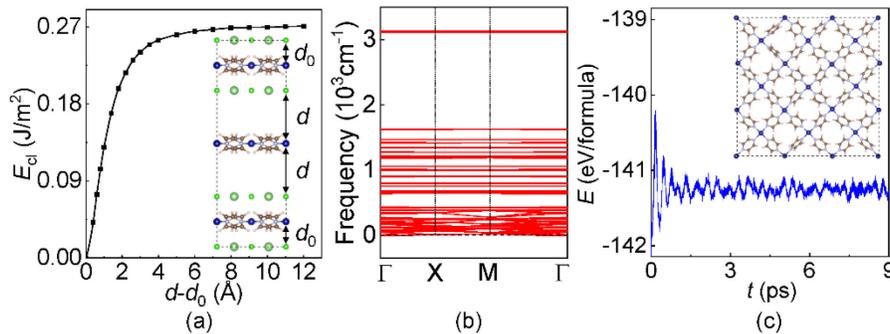

**Figure 2.** (a) Exfoliation energy $E_{cl}$ as a function of separation distance $d-d_0$ in the process of exfoliating one Cr(pyz)$_2$ layer from its bulk crystal. (b) Calculated phonon band structure of exfoliated Cr(pyz)$_2$ monolayer. (c) Total potential energy fluctuation of 3×3×1 supercell for Cr(pyz)$_2$ monolayer during AIMD simulations at 400 K. The inset shows the snapshot at the end of simulation of 9 ps.

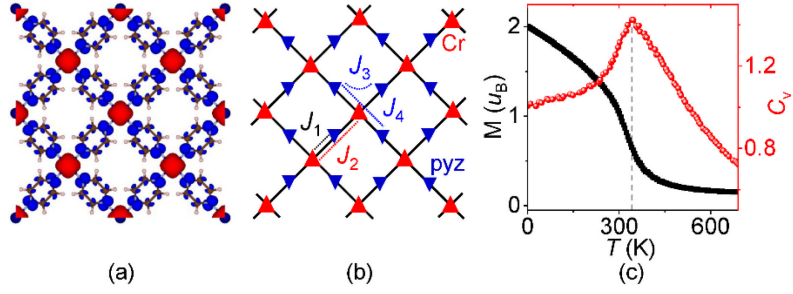

**Figure 3.** (a) Spin density distribution of the ground ferrimagnetic state for 2D Cr(pyz)$_2$ monolayer with an isovalue of 0.06 Å$^{-3}$. Red and blue indicate up and down spins, respectively. (b) Nearest-neighbor and next-nearest-neighbor spin exchange paths for Cr(pyz)$_2$ monolayer. The exchange-coupling parameter of $J_k$ ($k$ = 1~4) is also marked. $J_1$ represents the coupling between Cr and nearest-neighbor pyz. $J_2$/$J_3$ represents the coupling between the nearest two Cr/pyz. $J_4$ represents the coupling between next-nearest two pyz. (c) Magnetic moment (M) per formula (black) and specific heat $C_v$ (red) as a function of temperature from classic Heisenberg model Monte Carlo simulation.

To determine the magnetic ground state of Cr(pyz)$_2$ monolayer, five different magnetic states, namely one ferromagnetic (FM) state, one antiferromagnetic (AFM) state, and three ferrimagnetic (FiM) states (see Figure S1), are calculated to compare their relative energies. It is found that the structure prefers the FiM1 state in which the Cr spins are all antiparallelly aligned with the pyz spins. In the FiM1 state, the total magnetic moment is 2.0 $\mu_B$ per formula with a local magnetic moment of 3.4 $\mu_B$ per Cr and −0.6 $\mu_B$ per pyz. Accordingly, Cr and pyz possess a formal spin of 2 and 1/2, respectively. Figure 3(a) shows the spin density distribution in the ground FiM1 state. Obviously, the spin density on the pyz units is primarily contributed by the *p* orbitals of N atoms and only a small amount is distributed on the *p* orbitals of C atoms.

The ferrimagnetic coupling strength of Cr ions with pyz units is expected to be robust because of the large energy difference of 0.87 eV per formula between the FM and FiM1 states. To further estimate the Curie temperature $T_c$ of such ferrimagnetic coupling, we employ the Monte Carlo (MC) simulations based on the classic Heisenberg model Hamiltonian,[37]

$$H = -\sum_k \sum_{i>j} \sum_j J_k \times S_i \cdot S_j + \sum_i D_i S_{iz}^2 \quad (1)$$

where $J_k$ are the exchange-coupling parameters displayed in Figure 3(b). $D$ is the magnetic anisotropy parameter with a predicted value of 0.05 meV for Cr and 0 meV for pyz. The calculated values of $J_k$ are illustrated in Table S1. $S$ is the effective spin of Cr or pyz. Here, both the nearest-neighbor and next-nearest-neighbor spin exchange interactions are considered.

As shown in Figure 3(c), the simulated curve of spin magnetic moment (M) decreases from 2 $\mu_B$ to 0 with the increase of temperature. Correspondingly, the specific heat $C_v = (\langle E^2 \rangle - \langle E \rangle^2)/T^2$ is also calculated after the system reaches equilibrium at a given temperature. By locating the peak position of $C_v(T)$ plot, the ferrimagnetic Curie temperature $T_c$ is predicted to be 342 K, which is lower than that (510 K) of bulk crystal. Possible reasons are the neglected effective coupling between Cr(pyz)$_2$ layers, the unconsidered structural disorders present in experimentally synthesized bulk, and the underestimate trend of the classic Heisenberg model.

To reveal the electronic properties of 2D Cr(pyz)$_2$ monolayer, the band structures and density of states are calcu-lated with HSE06 functional. The Cr(pyz)$_2$ monolayer is found to be a direct semiconductor with a moderate band gap of 1.72 eV [Figure 4(a)]. Particularly, the valence band maximum (VBM) state is 100% spin-polarized in the spin-up channel, whereas the conduction band minimum (CBM) state is 100% spin-polarized in the spin-down channel; that is, it belongs to an intrinsic BMS.[25, 27, 30, 31] From the Kohn-Sham orbital distributions in Figure 4(b), one can see that the VBM state is dominated by the $p$ orbitals of N and C atoms, whereas the CBM state is built up by both the $p$ orbitals of N and C atoms and $3d$ orbitals of Cr ions, which is also clear from the projected density of states in Figure S2.

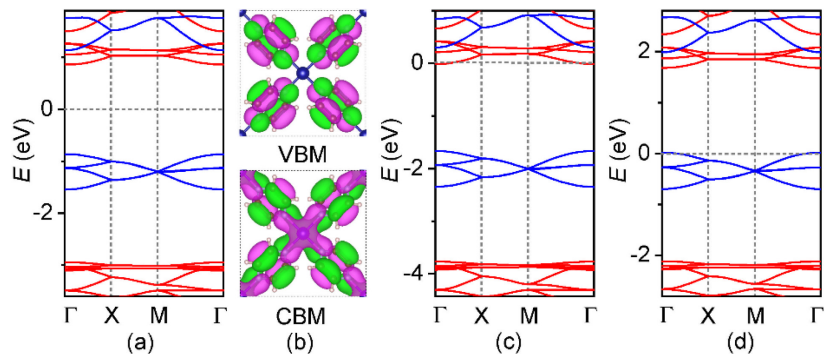

**Figure 4.** (a) Spin-polarized band structure of 2D Cr(pyz)$_2$ monolayer with HSE06 functional. Red and blue lines represent spin-up and spin-down bands, respectively. (b) Kohn-Sham orbitals of the valence band maximum (VBM) and conduction-band minimum (CBM) at Γ point with an isovalue of 1×10$^{-6}$ a. u.. Spin-polarized band structures for (c) electron doping and (d) hole doping with the carrier concentration of 1.1×10$^{13}$ cm$^{-2}$. Fermi levels are all set to zero.

The BMS electronic structure of Cr(pyz)$_2$ monolayer makes it promising for control of spin polarization by electrical gating. Here, to simulate the carrier doping effect induced by applied voltage gate, the electronic structures of electron and hole doped Cr(pyz)$_2$ monolayer, corresponding to positive and negative gate voltages respectively, are studied at the doping concentration of 0.05 e/h per formula or $1.1\times10^{13}$ cm$^{-2}$. As shown in Figures 4(c) and (d), under electron doping (positive gate voltage), the Fermi level shifts up into CBM, presenting half metallic conduction with a complete spin-down polarization, while the Fermi level shifts down into VBM under hole doping (negative gate voltage), half metallicity with a full spin-up polarization is obtained. This property allows us to manipulate the carriers' spin-polarization direction just by changing the sign of applied gate voltage.

## CONCLUSIONS

To summarize, based on experimentally synthesized Li$_{0.7}$[Cr(pyz)$_2$]Cl$_{0.7}$·0.25(THF), we propose to exfoliate Cr(pyz)$_2$ monolayer as a 2D magnetic semiconductor with room temperature ferrimagnetism and electrically controllable spin polarization. Such kind of organometallic ferrimagnetic semiconductors not only provide a new opportunity to achieve high-$T_c$ 2D magnetic semiconductors, but also has great potential in the design of electrically controlled nanospintronic devices.

## COMPUTATIONAL METHODS

First-principles calculations are carried out by using the density functional theory (DFT) method within the Perdew-Burke-Ernzerhof (PBE) generalized gradient approximation (GGA) implemented in Vienna ab initio Simulation Package (VASP).[38, 39] The strong-correlated correction is considered with GGA+U method[40] for structure optimization, phonon spectra calculation and ab initio molecular dynamic (AIMD) simulation. The values of effective onsite Coulomb interaction parameter ($U$) and exchange interaction parameter ($J$) are set as 3.0 and 1.0 eV, respectively, which predict the magnetic exchange energy for [Cr(pyz)$_4$]$^{2-}$ fragment close to that in the literature (see Figure S3).[15] Test calculations with different $U$ values give qualitatively the same conclusion (see Table S2). The DFT-D3 method with Becke-Jonson damping[41] is adopted for the van der Waals (vdW) interactions. The energy cutoff employed for plane-wave expansion of electron wave functions is set to be 520 eV. For sampling the first Brillouin zone, the Monkhorst-Pack $k$-point mesh is set with a grid spacing less than 0.02 Å$^{-1}$. During the structural optimization, the residual force on each atom is set to less than 0.01 eV/Å. The energy convergence is set to $1\times10^{-6}$ eV. A vacuum region of about 15 Å is applied along the z-direction to avoid mirror interaction between neighboring slabs for the simulation of Cr(pyz)$_2$ monolayer. The phonon spectra are calculated using the finite displacement method implemented in PHONOPY code together with the VASP code.[42] The thermal stability of the Cr(pyz)$_2$ monolayer is assessed according to the AIMD simulation at 400 K using a 3×3×1 supercell. The energy barrier from the C2h to D4h structure is investigated with the climbing image nudged elastic band (CI-NEB) method.[43] For accurately describe the band structure and density of states, we further apply the screened hybrid HSE06 functional,[44, 45] which includes the accurate Fock exchange and usually performs much better than the GGA and GGA+U methods.[46-48]

## ASSOCIATED CONTENT

**Supporting Information**. Additional material includes spin density distribution of different magnetic states for Cr(pyz)$_2$ monolayer and [Cr(pyz)$_4$]$^{2-}$ fragment; projected density of states and magnetic exchange parameters for Cr(pyz)$_2$ monolayer; relative energies of FM to AFM state for [Cr(pyz)$_4$]$^{2-}$ fragment at different $U$ values. This material is available free of charge via the Internet at http://pubs.acs.org.

## AUTHOR INFORMATION

**Corresponding Author**

* xjwu@ustc.edu.cn
* lixx@ustc.edu.cn
* jlyang@ustc.edu.cn
△ X.Y.L. and H.F.L. contributed equally to this work.
**Notes**
The authors declare no competing financial interest.

## ACKNOWLEDGMENT

This work is supported by the National Natural Science Foundation of China (Grant No. 21688102), by the National Key Research & Development Program of China (Grant No. 2016YFA0200604), by Anhui Initiative in Quantum Information Technologies (Grant No. AHY090400), by the Youth Innovation Promotion Association CAS (2019441), and by USTC Research Funds of the Double First-Class Initiative (YD2060002011). The computational resources are provided by the Supercomputing Center of University of Science and Technology of China, Supercomputing Center of Chinese Academy of Sciences, and Tianjin and Shanghai Supercomputer Centers.

Table of Contents artwork

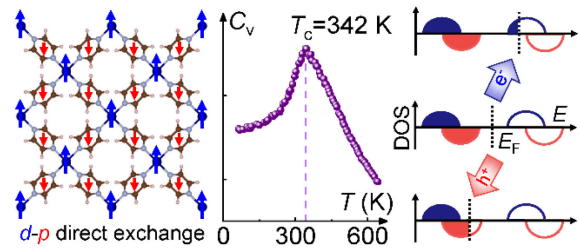

Supporting Information

# Two-Dimensional Bipolar Magnetic Semiconductor with High Curie Temperature and Electrically Controllable Spin Polarization Realized in Exfoliated Cr(pyrazine)$_2$ Monolayer

Xiangyang Li,[†,§,Δ] Haifeng Lv,[†,§,Δ] Xiaofeng Liu,[†,§] Xiaojun Wu,*[†,§] Xingxing Li,*[†,§] and Jinlong Yang*[†,‡,§]

†Synergetic Innovation Center of Quantum Information and Quantum Physics, University of Science and Technology of China, Hefei, Anhui 230026, China

‡Hefei National Laboratory for Physical Sciences at the Microscale, University of Science and Technology of China, Hefei, Anhui 230026, China

§Department of Chemical Physics, University of Science and Technology of China, Hefei, Anhui 230026, China

*E-mail: xjwu@ustc.edu.cn

*E-mail: lixx@ustc.edu.cn

*E-mail: jlyang@ustc.edu.cn

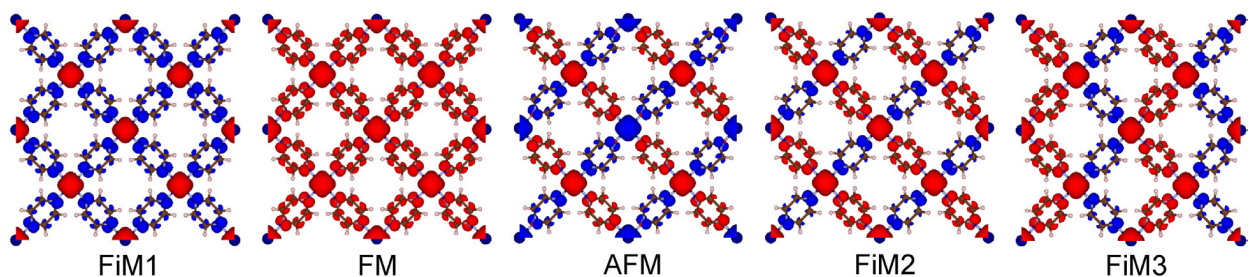

**Figure S1.** Spin density distribution of five different magnetic states for Cr(pyz)$_2$ monolayer, namely one ferromagnetic (FM) state, one antiferromagnetic (AFM) state, and three ferrimagnetic (FiM) states, with an isovalue of 0.06 Å$^{-3}$.

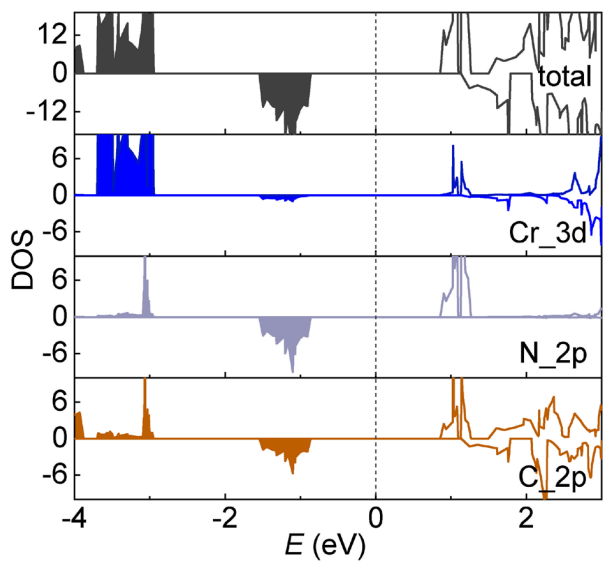

**Figure S2.** Projected density of states (DOS) for 2D Cr(pyz)$_2$ monolayer with HSE06 functional. The Fermi level is set to zero.

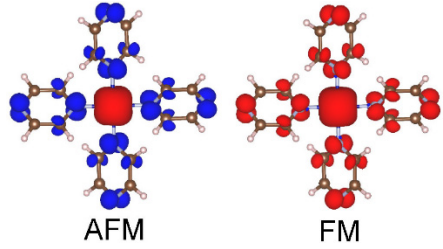
AFM    FM

**Figure S3.** Spin density distribution of AFM and FM states for $[Cr(pyz)_4]^{2-}$ fragment with an isovalue of 0.06 Å$^{-3}$. When the onsite Coulomb interaction $U$ is set as 3 eV, the calculated relative energy of FM configuration to AFM configuration is 0.84 eV, which is basically the same as that of 0.82 eV in the literature.[1]

**Table S1.** The magnetic energies (eV per formula) of four different magnetic configurations relative to FiM1 configuration, and deduced magnetic exchange parameters $J_k$ (meV) for Cr(pyz)$_2$ monolayer.

|  | Cr(pyz)$_2$ |  | Cr(pyz)$_2$ |
|---|---|---|---|
| $E_{FM}$ | 0.868 | $J_1$ | 108.4 |
| $E_{AFM}$ | 0.350 | $J_2$ | 2.7 |
| $E_{FiM2}$ | 0.393 | $J_3$ | 20.3 |
| $E_{FiM3}$ | 0.406 | $J_4$ | 7.3 |

**Table S2.** Calculated relative energies $\Delta E$ (eV per formula) of FM configuration to AFM configuration for $[Cr(pyz)_4]^{2-}$ fragment at different values of onsite Coulomb interaction $U$ (eV).

| $[Cr(pyz)_4]^{2-}$ | $U=2$ | $U=3$ | $U=4$ | $U=5$ | $U=6$ | Ref. [1] |
|---|---|---|---|---|---|---|
| $\Delta E$ | 1.051 | 0.841 | 0.690 | 0.579 | 0.493 | 0.816 |